\documentstyle[12pt]{article}

\def\hybrid{\topmargin -20pt	\oddsidemargin 0pt
	\headheight 0pt	\headsep 0pt
	\textwidth 6.25in	
	\textheight 9.5in	
	\marginparwidth .875in
	\parskip 5pt plus 1pt	\jot = 1.5ex}

\hybrid

\def\baselinestretch{1.2}

\catcode`\@=11

\def\marginnote#1{}
\newcount\hour
\newcount\minute
\newtoks\amorpm
\hour=\time\divide\hour by60
\minute=\time{\multiply\hour by60 \global\advance\minute by-\hour}
\edef\standardtime{{\ifnum\hour<12 \global\amorpm={am}%
	\else\global\amorpm={pm}\advance\hour by-12 \fi
	\ifnum\hour=0 \hour=12 \fi
	\number\hour:\ifnum\minute<10 0\fi\number\minute\the\amorpm}}
\edef\militarytime{\number\hour:\ifnum\minute<10 0\fi\number\minute}

\def\draftlabel#1{{\@bsphack\if@filesw {\let\thepage\relax
   \xdef\@gtempa{\write\@auxout{\string
      \newlabel{#1}{{\@currentlabel}{\thepage}}}}}\@gtempa
   \if@nobreak \ifvmode\nobreak\fi\fi\fi\@esphack}
	\gdef\@eqnlabel{#1}}
\def\@eqnlabel{}
\def\@vacuum{}
\def\draftmarginnote#1{\marginpar{\raggedright\scriptsize\tt#1}}

\def\draft{\oddsidemargin -.5truein
	\def\@oddfoot{\sl preliminary draft \hfil
	\rm\thepage\hfil\sl\today\quad\militarytime}
	\let\@evenfoot\@oddfoot	\overfullrule 3pt
	\let\label=\draftlabel
	\let\marginnote=\draftmarginnote
   \def\@eqnnum{(\theequation)\rlap{\kern\marginparsep\tt\@eqnlabel}%
\global\let\@eqnlabel\@vacuum}  }

\def\preprint{\twocolumn\sloppy\flushbottom\parindent 2em
	\leftmargini 2em\leftmarginv .5em\leftmarginvi .5em
	\oddsidemargin -.5in	\evensidemargin -.5in
	\columnsep .4in	\footheight 0pt
	\textwidth 10.in	\topmargin  -.4in
	\headheight 12pt \topskip .4in
	\textheight 6.9in \footskip 0pt
	\def\@oddhead{\thepage\hfil\addtocounter{page}{1}\thepage}
	\let\@evenhead\@oddhead	\def\@oddfoot{}	\def\@evenfoot{} }

\def\numberbysection{\@addtoreset{equation}{section}
	\def\theequation{\thesection.\arabic{equation}}}

\def\underline#1{\relax\ifmmode\@@underline#1\else
	$\@@underline{\hbox{#1}}$\relax\fi}

\def\titlepage{\@restonecolfalse\if@twocolumn\@restonecoltrue\onecolumn
     \else \newpage \fi \thispagestyle{empty}\c@page\z@
	\def\thefootnote{\fnsymbol{footnote}} }

\def\endtitlepage{\if@restonecol\twocolumn \else \newpage \fi
	\def\thefootnote{\arabic{footnote}}
	\setcounter{footnote}{0}}  

\catcode`@=12
\relax

\def\figcap{\section*{Figure Captions\markboth
	{FIGURECAPTIONS}{FIGURECAPTIONS}}\list
	{Figure \arabic{enumi}:\hfill}{\settowidth\labelwidth{Figure
999:}
	\leftmargin\labelwidth
	\advance\leftmargin\labelsep\usecounter{enumi}}}
 \relax
\def\tablecap{\section*{Table Captions\markboth
	{TABLECAPTIONS}{TABLECAPTIONS}}\list
	{Table \arabic{enumi}:\hfill}{\settowidth\labelwidth{Table
999:}
	\leftmargin\labelwidth
	\advance\leftmargin\labelsep\usecounter{enumi}}}
 \relax
\def\reflist{\section*{References\markboth
	{REFLIST}{REFLIST}}\list
	{[\arabic{enumi}]\hfill}{\settowidth\labelwidth{[999]}
	\leftmargin\labelwidth
	\advance\leftmargin\labelsep\usecounter{enumi}}}
 \relax
\makeatletter
\newcounter{pubctr}
\def\publist{\@ifnextchar[{\@publist}{\@@publist}}
\def\@publist[#1]{\list
	{[\arabic{pubctr}]\hfill}{\settowidth\labelwidth{[999]}
	\leftmargin\labelwidth
	\advance\leftmargin\labelsep
	\@nmbrlisttrue\def\@listctr{pubctr}
	\setcounter{pubctr}{#1}\addtocounter{pubctr}{-1}}}
\def\@@publist{\list
	{[\arabic{pubctr}]\hfill}{\settowidth\labelwidth{[999]}
	\leftmargin\labelwidth
	\advance\leftmargin\labelsep
	\@nmbrlisttrue\def\@listctr{pubctr}}}
 \relax
\makeatother
\newskip\humongous \humongous=0pt plus 1000pt minus 1000pt

\newif\ifdtup

\relax

\def\be{\begin{equation}}
\def\ee{\end{equation}}
\def\ba{\begin{eqnarray}}
\def\ea{\end{eqnarray}}

\def\del{\partial}

\def\r{\rho}

\def\d{\delta}

\def\e{\epsilon}

\def\th{\theta}

\def\m{\mu}
\def\n{\nu}

\def\s{\sigma}

\def\cg{{\cal{G} }}
\def\cD{{\cal{D} } }

\def\no{\noindent}

\def\IR{\relax{\rm I\kern-.18em R}}

\def \ha {{1\over 2}}

\def \ov {\over}

\def\IR{\relax{\rm I\kern-.18em R}}
\def\inv{^{\raise.15ex\hbox{${\scriptscriptstyle -}$}\kern-.05em 1}}

\def\PL{Poisson--Lie T--duality}

\begin{document}

\renewcommand{\theequation}{\arabic{equation}}
\newcommand{\beq}{\begin{equation}}
\newcommand{\eeq}[1]{\label{#1}\end{equation}}
\newcommand{\ber}{\begin{eqnarray}}
\newcommand{\eer}[1]{\label{#1}\end{eqnarray}}
\newcommand{\eqn}[1]{(\ref{#1})}
\begin{titlepage}
\begin{center}

\hfill THU--96/38\\
\hfill September 1996\\
\hfill hep--th/9611199\\

\vskip .8in

{\large \bf POISSON--LIE T--DUALITY AND SUPERSYMMETRY
\footnote{Contribution to the proceedings of the 
{\em 30th International Symposium Ahrenshoop on the Theory of Elementary
Particles}, Buckow, Germany, 26-31 August 1996 (invited talk).}}

\vskip 0.6in

{\bf Konstadinos Sfetsos
\footnote{e--mail address: sfetsos@fys.ruu.nl}}\\
\vskip .1in

{\em Institute for Theoretical Physics, Utrecht University\\
     Princetonplein 5, TA 3508, The Netherlands
\footnote{Presently serving in the Greek Armed Forces.}}\\

\vskip .2in

\end{center}

\vskip .6in

\begin{center} {\bf ABSTRACT } \end{center}
\begin{quotation}\noindent

\no
We review aspects of Poisson--Lie T--duality which we explicitly 
formulate as a canonical transformation on the world--sheet. 
Extensions of previous work on T--duality
in relation to supersymmetry are also discussed.

\vskip .2in

\noindent

\end{quotation}

\end{titlepage}

\def\baselinestretch{1.2}
\baselineskip 16 pt
\noindent


\section{Introduction}

Duality symmetries in string theory reveal phenomena 
with no field theoretical analogs. As such they 
should be used in order to resolve long standing problems in
fundamental Physics. 
Unlike the inherently non--perturbative S--duality and the dualities 
between apparently  different string theories, T--duality \cite{BUSCHER}
is rather well understood. 
Its interplay with supersymmetry served as a useful tool 
in discovering such phenomena,
since in certain cases Abelian as well as non--Abelian 
T--duality are in conflict with supersymmetry from a field 
theoretical point of view  \cite{bakasII,KSsusynab} but in 
complete harmony in a string theoretical setting \cite{basfe,KSsusynab} 
(see also \cite{hassand,AAB}). T--duality as a supersymmetry 
breaking--restoration mechanism has been discussed in \cite{sferesto}
in connection also with revised versions of some key theorems in 
2-dim $\s$--models with extended supersymmetry \cite{zumino}.
All these should have implications, yet unexplored, for 
supersymmetry breaking scenarios in string phenomenology and they might
hint the mechanism that resolves, in a string theoretical 
framework, the information loss paradox in black hole Physics.

In order to be able to compare with more general cases to be discussed in 
the rest of the paper we will review some results concerning 
world--sheet supersymmetry and T--duality leaving aside the case of 
target space supersymmetry where phenomena of similar origin also occur
\cite{bakasII,BKO,hassand,AAB,sferesto}. 
The case of $N=1$ supersymmetry presents no problem
from the point of view of its behaviour under T--duality since both 
dual backgrounds can be made $N=1$ supersymmetric \cite{zasymp,hassand}. 
Interesting situations
arise only when we consider $N=2$ and especially $N=4$ extended
supersymmetry. The conventional \cite{zumino}
definition of 
N--extended world--sheet supersymmetry implies that there exist $N-1$
complex structures $F_I^\pm$, $I=1,2,\dots , N-1$
in the two chiral sectors separately which are covariantly
constant.
It is useful to assign the complex structures into representations 
of the isometry group $G$ with respect to
which the duality is performed. Let's denote by $\{v^a\}$ the set of
vector fields generating the algebra for $G$
and by $\pounds_{v^a}$ the corresponding 
Lie--derivative. We concentrate on the case of $N=4$ world--sheet 
supersymmetry and distinguish the 
following interesting cases \cite{basfe,KSsusynab}:

\no
\underline{case (i)}: The duality group is $G\simeq U(1)$ and all
three complex structures are singlets. Then the corresponding Abelian 
T--duality preserves $N=4$.
Symbolically we write
\be
\left\{ {\pounds_v} F^\pm_I = 0~ , ~~~  I=1,2,3\right \}
~~~~\stackrel{duality}{\Longrightarrow} ~~~~  N=4 ~ .
\label{ca1}
\ee

\no
\underline{case (ii)}: The duality group is $G\simeq U(1)$ but only
one complex structure is a singlet, whereas the other two 
form a doublet. Then the 
Abelian T--duality preserves the part of the 
original $N=4$ generated by the complex structure 
which is a singlet. Hence we have
\be
\left\{ \begin{array}{ccc} 
{\pounds_v} F^\pm_3 &  = & 0 \\
{\pounds_v} F^\pm_1 & = & F^\pm_2 \\
{\pounds_v} F^\pm_2 & = & - F^\pm_1
\end{array} 
\right \}
~~~~\stackrel{duality}{\Longrightarrow}~~~~ N=2 ~ .
\label{ca2}
\ee
The rest of supersymmetry
is realized non--locally with corresponding complex structures that 
are non--local functionals of the dual target space variables. 
Hence, although supersymmetry seems to be broken after T--duality
from the low energy field theory point of view, 
it is nevertheless restored in a string theoretical framework.

\no
\underline{case (iii)}: The duality group is $G\simeq SO(3)$ and all
three complex structures are singlets. Then the corresponding 
non--Abelian T--duality preserves $N=4$, namely 
\be
\left\{ {\pounds_{v^a}} F^\pm_b = 0~ , ~~~  a,b=1,2,3\right \}
~~~~\stackrel{duality}{\Longrightarrow}~~~~ N=4 ~ .
\label{ca3}
\ee

\no
\underline{case (iv)}: The duality group is $G\simeq SO(3)$ but
the complex structures belong to the triplet representation of it. Then the
non--Abelian T--duality breaks completely the extended supersymmetry
from the low energy effective field theory point of view. Symbolically
\be
\left\{ {\pounds_{v^a}} F^\pm_b  =  \e_{abc} F^\pm_c ~ ,~~~ 
a,b,c=1,2,3\right \} ~~~~\stackrel{duality}{\Longrightarrow}
~~~~  N=1 ~ .
\label{ca4}
\ee
Needless to say that supersymmetry is restored in full at the string level 
and is realized with non--local complex structures.
Prototype examples realizing all of the above cases 
have been explicitly worked out \cite{basfe,KSsusynab} and
are hyper--kahler metrics with translational or rotational groups of
isometries, which include
the Eguchi--Hanson, the Taub--NUT and the Atiyah--Hitchin
metrics. 

In addition it has been shown \cite{basfe,sferesto,KSsusynab}
that in some cases where the exact 
conformal field theory corresponding to the dual backgrounds is known, 
the non--local realizations of supersymmetry after duality are naturally
represented using classical parafermions \cite{BCR}. 
Hence, direct contact with corresponding realizations of the $N=4$ 
superconformal algebra \cite{KAFK} has been made.
The most elementary example for both
Abelian and non--Abelian duality is the background corresponding
to the WZW model for $SU(2) \otimes U(1)$.
Finally, let us mention that all theorems which have been proved in the 
past concerning extended world--sheet 
supersymmetry, implicitly assumed local realizations
of it \cite{zumino} and  
they are not valid when it is non--locally realized. 
For instance, it is not always true that when the string torsion vanishes
$N=4$ extended supersymmetry implies that the manifold is Ricci flat.
The interesting reader will find revised versions of some of these
theorems, including the one we just mentioned,
in \cite{sferesto}. Here we only mention that in general 
one can show that when the complex 
structures are non--local functionals of the target space variables
they do not have to be covariantly constant in order to define an 
extended supersymmetry but they should obey instead \cite{sferesto}
\be
D^\pm_\m F^\pm_{\n\r} \del_\mp X^\m + \tilde \del_\mp F^\pm_{\n\r} = 0~ ,
\label{nocos}
\ee
where the tilded world--sheet derivative acts only on the non--local 
part of the complex structure. 

In order to further improve our picture 
it remains to examine in this context 
the so called \PL\ \cite{KliSevI}, which
generalizes the concept of T--duality as it can be performed 
even in the absence of isometries. Hence, it is an advancement both 
conceptually as well as in practice since isometries are not always preserved 
under T--duality anyway.
In this paper we address 
issues involved in the interplay between \PL\ and supersymmetry.
In particular, we find the modified version
of \eqn{ca1}--\eqn{ca4} and remark on the potential 
physical applications.

The organization of this paper is as follows: In section 2 we review 
aspects of \PL\ \cite{KliSevI,DriDou} which are directly relevant 
in our context and 
in addition we explicitly formulate it as a canonical transformation 
on the world--sheet.
In this respect we put \PL\ in equal footing
with T--duality in the presence of isometries
as the latter has already been given an analogous 
formulation \cite{AALcan,zacloz,KSsusynab}.
In section 3 we start examining supersymmetric \PL\ by 
presenting its on shell formulation in $\s$--models with $N=1$
as well as extended world--sheet supersymmetry.
We end the paper in section 4 with concluding remarks and discussion 
on the future
directions of this work as well as on its relevance 
in resolving some fundamental problems in Physics.

\section{Poisson--Lie T--duality}

We consider classical closed string propagation in 
$d$-dimensional backgrounds. 
Some of the 
target space variables are chosen to parametrize an element 
$g$ of a group $G$ (with algebra $\cg$) and  
will be denoted by $X^\m$, $\m=1,2,\dots  ,dim(G)$, whereas 
the rest are the so called spectators and will be denoted by
$Y^i,i=1,2,\dots, d-dim(G)$.
We also introduce representation matrices
$\{T_a\}$, with $a=1,2,\dots, dim(G)$ and
the components of the left--invariant Maurer--Cartan forms 
$ L^a_\pm = L^a_\m \del_\pm X^\m$. 
The inverse of $L^a_\m$ will be denoted by $L^\m_a$ 
and for notational convenience we will also use $L^i_\pm \equiv \del_\pm Y^i$.
Then the most general $\s$--model action is given by
(the light--cone coordinates on the world--sheet are
$\s^\pm =\ha (\tau \pm \s)$)
\ba S & = & \ha \int L^A_+ E^+_{AB} L^B_- 
\nonumber \\
&  = & \ha \int L_+^a E^+_{ab} L_-^b + \del_+ Y^i \Phi^+_{ia} L_-^a 
+ L_+^a \Phi^+_{ai} \del_- Y^i + \del_+ Y^i \Phi_{ij} \del_- Y^j~ ,
\label{smoac}
\ea
where the index $A=\{a,i\}$.
The couplings $E^+_{ab}$, $\Phi^+_{ia}$, $ \Phi^+_{ai}$ 
(for later use we also define $E^-_{ab}=E^+_{ba}$, $\Phi^-_{ia}=
\Phi^+_{ai}$ and $\Phi^-_{ai}=\Phi^+_{ia}$)
and $ \Phi_{ij}$ may depend on all variables $X^\m$ and $Y^i$. 
Hence, we do not require any isometry associated 
with the group $G$.

Another $\s$--model (denoted as usual with
tilded symbols) is said to be dual to \eqn{smoac}
in the sense of Poisson--Lie T--duality \cite{KliSevI}
if the algebras $\cg$ and $\tilde \cg$ form a pair 
of maximally isotropic subalgebras the Lie algebra $\cD$ of a 
Lie group $D$ known as the Drinfeld double can be decomposed to.
Leaving aside the mathematical details
(see for instance \cite{alemal}), this 
implies the non--trivial commutator
\be
[T_a,\tilde T^b ] = 
i \tilde f^{bc}{}_a T_c - i f_{ac}{}^b \tilde T^c ~ .
\label{mixbra}
\ee
The Jacobi identity $[[T_a,T_b],\tilde T^c] + {\rm cyclic} = 0$ 
relates the structure constants of the two algebras as \cite{alemal,KliSevI}
\be
f_{ab}{}^d \tilde f^{ce}{}_d + f_{d[a}{}^c \tilde f^{de}{}_{b]}
- f_{d[a}{}^e \tilde f^{dc}{}_{b]} = 0 ~ .
\label{ftilf}
\ee
This restricts severely the candidate algebras 
to form Drinfeld doubles. 
Besides the trivial solution to \eqn{ftilf},
when the group $G$ or $\tilde G$ is Abelian, 
other non--trivial solutions 
exist and typically involve the decomposition of certain semi--simple
Lie groups into factors that contain Borel subgroups \cite{alemal}. 
In addition to \eqn{mixbra}, there is a bilinear invariant 
$\langle{\cdot|\cdot \rangle}$ with the various generators obeying 
\be
\langle{T_a|T_b\rangle}= \langle{\tilde T^a|\tilde T^b \rangle}= 0~ ,
~~~~ \langle{T_a|\tilde T^b \rangle} = \d_a{}^b ~ .
\label{bili}
\ee
We also define matrices $a(g)$, $b(g)$ and $\Pi(g)$ as
\be
 g\inv T_a g = a_a{}^b T_b~ ,~~~~ g\inv \tilde T^a g = 
b^{ab} T_b +  (a\inv)_b{}^a \tilde T^b~ ,~~~~ \Pi^{ab}=
b^{ca} a_c{}^b ~ .
\label{abpi}
\ee
Consistency then requires that 
\be
a(g\inv) = a\inv(g)~ ,~~~~ b^T(g)= b(g\inv)~ ,~~~~
\Pi^T(g) = - \Pi(g) ~ .  
\label{conss}
\ee 
Then the various couplings in the $\s$--model action \eqn{smoac} 
are restricted to be of the form \cite{KliSevI,TyuUng}
(although occasionally suppressed
in the rest of the paper, the index structure should be clear)
\ba 
&& \Phi^\pm = E^\pm (E_0^\pm)\inv F^\pm ~ ,~~~~ 
E^\pm = \left( (E_0^\pm)\inv \pm \Pi\right)\inv ~ ,\nonumber \\ 
&& \Phi = F -  F^+ \Pi E^+ (E^+_0)\inv F^+ ~ ,
\label{coupl}   
\ea
where the new couplings 
$F^+_{ia}=F^-_{ai}$, $F^+_{ai}=F^-_{ia}$ and $(E_0^+)_{ab}= (E_0^-)_{ba}$ 
maybe functions of the 
spectator variables $Y^i$ only.
The couplings of the dual action are also determined in a similar 
fashion \cite{KliSevI,TyuUng}
\ba 
&& \tilde \Phi^\pm = \pm \tilde E^\pm F^\pm ~ ,~~~~~
\tilde E^\pm = \left(E_0^\pm \pm \tilde \Pi\right)\inv ~ , \nonumber \\ 
&& \tilde \Phi = F - F^+ \tilde E F^+ ~ .
 \label{coupldu}
\ea
The traditional non--Abelian duality is recoved if $\tilde G$ 
(equivalently $G$) is Abelian. 
Then it is easy to see using \eqn{abpi} that 
\be
b^{ab}=\Pi^{ab}=0~ ,~~~ \tilde a^a{}_b = \d^a{}_b ~ ,~~~
\tilde b_{ab} = f_{ab}{}^c \chi_c = -\tilde \Pi_{ab}~ ,~~~
(\tilde L_\pm)_a = \del_\pm \chi_a~ ,
\label{abadou}
\ee 
where the $\chi_a$'s parametrize the group element
$\tilde g= e^{i \chi_a \tilde T^a} \in \tilde G$.
Then action \eqn{smoac}
becomes invariant under left transformations of $g\in G$, i.e. there is an
isometry and the dual action 
reduces to the one computed in \cite{grnonab}.

\subsection*{Poisson--Lie T--duality as a canonical transformation}

\PL\ is by definition a canonical transformation and an expression for
the generating functional has been given \cite{DriDou}. Here we 
formulate \PL\ as an explicit   
transformation between the canonical variables of 
the two dual $\s$--models. 
This was done for the case of Abelian duality in \cite{AALcan},
for non--Abelian duality on Principal Chiral models
in \cite{zacloz} and for the 
general $\s$--model \eqn{smoac} when it is invariant under the 
left (or right) action of the group $G$, i.e. when \eqn{abadou} holds,
in \cite{KSsusynab}.
It is instructive to see 
how this can be used to make an educated guess for the 
corresponding canonical 
transformation for Poisson--Lie T--duality.
We recall that when \eqn{smoac} is invariant under the left action 
of the group $G$ the transformation between the canonical pairs of
variables $(L^a_\s, P_b)$ and $(\del_\s \chi_a, \tilde P^b)$ is given by
\ba
L_\s^a & = & \tilde P^a ~ ,
\label{catra1} \\
P_a & = & - f_{ab}{}^c \chi_c \tilde P^b + \del_\s \chi_a ~ .
\label{catra2}
\ea
where $L^a_\s \equiv L^a_\m \del_\s X^\m$, $  P_a \equiv  L^\m_a P_\m $ 
and similarly for $(\tilde L_\s)_a$ and $\tilde P^a$.
Rewriting \eqn{catra2} with the help of \eqn{abadou} suggests the following 
ansatz for a canonical transformation corresponding to Poisson--Lie T--duality
\ba
L^a_\s & = & (\d^a{}_b - \Pi^{ac} \tilde \Pi_{cb}) \tilde P^b - 
\Pi^{ab} (\tilde L_\s)_b ~ , \label{cantr1}  \\
P_a & = & \tilde \Pi_{ab} \tilde P^b +  (\tilde L_\s)_a ~ .
\label{cantr}
\ea
This ansatz is duality invariant since it implies a 
transformation similar to \eqn{cantr} with tilded and 
untilded symbols exchanged.\footnote{It is also manifest when we
cast \eqn{cantr1}, \eqn{cantr} into the form
\ba
P & = & ( I -\tilde \Pi \Pi)\inv (\tilde L_\s + \tilde \Pi L_\s) ~ , 
\nonumber \\
\tilde P & = & ( I -\Pi \tilde \Pi)\inv ( L_\s +  \Pi \tilde L_\s) ~ .
\label{ptp}
\ea
}
It is important to check whether or not \eqn{cantr1}, \eqn{cantr}
indeed constitute a canonical transformation since there are obviously
infinitely many transformations that reduce to \eqn{catra1}, \eqn{catra2}
when \eqn{abadou} holds.
Notice that the right hand side of
\eqn{cantr1} depends on tilded as well as untilded variables. 
Hence, it is not immediately  
obvious that the $\s$--model arising after sucsh a transformation will be
well defined, i.e. local in the dual target space variables. 
For simplicity we will not consider spectator fields,
since in any case they enter the canonical 
transformation implicitly via the definitions of $P_a$
and $\tilde P^a$. 
Then, the Hamiltonian associated with the action \eqn{smoac} 
is (we omit writing the $\s$--integration)
\be
H = \ha (P_a - B_{ac} L^c_\s) G^{ab} (P_b - B_{bd} L^d_\s) + 
\ha L^a_\s G_{ab} L^b_\s  ~ ,
\label{hamil}
\ee
where $G_{ab}$ and $B_{ab}$ are the symmetric and antisymmetric parts
of $E^+_{ab}$ 
\ba
 G & = &
 (I + E_0^+ \Pi)\inv G_0 (I - \Pi E_0^-)\inv ~ ,
\nonumber \\
 B &  =  &
(I + E_0^+ \Pi)\inv (B_0 - E^+_0 \Pi E^-_0) 
(I - \Pi E_0^-)\inv ~ .
\label{GB}
\ea
It is a straightforward, although a technically laborious,
computation to verify that the transformation \eqn{cantr1}, \eqn{cantr}
gives the Hamiltonian corresponding to the dual to  \eqn{smoac} action.
Hence, it is defined entirely 
in the dual model and potential non--localities
associated with the particular form of \eqn{cantr1}, 
have not arised.
It also turns out that \eqn{cantr1}, \eqn{cantr} are valid even
when we turn on the spectator fields which are then left invariant 
under the canonical transformation. 
In principle in order for \eqn{cantr1}, \eqn{cantr} to be a canonical 
transformation they have, in addition to generating the correct Hamiltonian
for the dual model,
to preserve the canonical Poisson brackets for the conjugate pair 
of variables $\{L^a_\m, P_a\}$ (see for instance \cite{KSsusynab}).
This is equivalent to finding a generating functional for 
the transformation. As already mentioned an 
implicit expression for such a
generating functional has been given in \cite{DriDou}. 
It is important to check 
whether or not it indeed reproduces our explicit transformation \eqn{cantr1}, 
\eqn{cantr}.

\subsection*{Poisson--Lie T--duality at the Drinfeld double}

Next we present a manifest formulation of Poisson--Lie T--duality directly
at the level of the Drinfeld double.
We will basically follow \cite{DriDou} but we will also include the
spectator fields in our discussion explicitly. At the end of this subsection
we will comment on an alternative formulation \cite{TyuUng}.

We introduce a basis of vectors $R^\pm_a$ satisfying
\be
\langle{R^\pm_a|R^\pm_b\rangle} = \pm \eta_{ab}~ , ~~~~ 
\langle{R^+_a|R^-_b\rangle} = 0 ~ ,
\label{ortth}
\ee
and the completeness relation
\be 
|R^+_a {\rangle} \eta^{ab} \langle{ }R^+_b| 
- |R^-_a {\rangle} \eta^{ab} \langle{ }R^-_b|   = I ~ , 
\label{Rcopl} 
\ee  
where $\eta_{ab}$ and $\eta^{ab}$ are some metric and its inverse.
We also introduce an operator $R$ as 
\be
R = |R^+_a {\rangle} \eta^{ab} \langle{ }R^+_b| 
+ |R^-_a {\rangle} \eta^{ab} \langle{ }R^-_b|  ~ . 
\label{RRpRm}
\ee
A useful representation 
in terms of $T_a$ and $\tilde T^a$ is 
\be
R^\pm_a  =  T_a \pm (E_0^\pm)_{ab} \tilde T^b ~ ,
~~~~ \eta_{ab}= (E_0^+)_{ab} + (E_0^-)_{ab} ~ .
\label{rree}
\ee
Then consider an action defined in the Drinfeld double as
\ba
 S & = & I_{0}(l) + {1\ov 2\pi} 
\int \langle{l\inv \del_\s l |R| l\inv \del_\s l \rangle} 
\nonumber  \\
&& 
- 2 i \del_+ Y F^+ \eta\inv \langle{l\inv \del_\s l|R^-\rangle}
-2 i \langle{l\inv \del_\s l | R_+ \rangle} \eta\inv F^+ \del_- Y
\nonumber \\
&& 
+ \del_+ Y F \del_- Y -\ha (\del_+ Y F^+ + \del_- Y F^-)
\eta\inv (F^- \del_+ Y + F^+ \del_- Y ) ~ .
\label{actiL}
\ea
The first line is the action introduced in \cite{DriDou}, where
$I_{0}(l)$ is the WZW action for a group element $l\in D$
with the world--sheet 
variables $\s$ and $\tau$ playing the role of light--cone variables.
The second line contains natural coupling terms 
between the spectator fields and
the 1--form in the double $l\inv\del_\s l$. The third line depends purely on
spectator fields. With the help of the Polyakov--Wiegman formula 
is easily seen that \eqn{actiL} is invariant 
under $l(\tau,\s)\to l_0(\tau) l(\tau,\s)$ for some $\tau$--dependent
element $l_0\in D$.
This local invariance can be used to cast the equations of 
motion for \eqn{actiL} with respect to variations of $l$ into the
form
\be
\langle{l\inv \del_\pm l |R^\mp_a\rangle}  = \pm i \del_\pm Y^i F^\pm_{ia}~ . 
\label{exil}
\ee
In the vicinity of the unit element of $D$ we may decompose \cite{alemal} the 
group element $l\in D$ as $l=\tilde h g$ or as $l=h \tilde g$, where
$h,g \in G$ and $\tilde h, \tilde g \in \tilde G$.
Using the first decomposition we may solve for the ``gauge field''
$A_\pm=\tilde h\inv \del_\pm \tilde h \in \tilde \cg$ with result
\ba
A_\pm & = &\left(\pm i \del_\pm Y F^\pm - \langle{g\inv\del_\pm g|R^\mp\rangle}
\right) M_\mp\inv~ ,~~~~ M_\pm = \langle{g\inv \tilde T g|R^\pm \rangle} 
\nonumber \\
& = &  \pm i a E^\mp \left( L_\pm + (E^\mp_0)\inv F^\mp \del_\pm Y \right)~ ,
\label{apam}
\ea
where in order to write the second line we have used \eqn{rree}.
Using the second decomposition it is clear that we may solve for 
$\tilde A_\pm= h\inv \del_\pm h \in  \cg$ with result 
\be
\tilde A_\pm = \pm i \tilde a \tilde E^\mp 
\left(\tilde L_\pm \mp F^\mp \del_\pm Y \right) ~ .
\label{appam}
\ee
We also note that the vanishing of the curvature associated with $A_\pm$
or $\tilde A_\pm$ 
boils down to condition \eqn{ftilf} on the
structure constants of the bialgebra $\cD$.
In order to recover \eqn{smoac} with couplings \eqn{coupl}
we insert the decomposition $l=\tilde h g$ into \eqn{actiL} and we use the
Polyakov--Wiegman formula and \eqn{bili}. It turns out that the 
action depends quadratically on 
$A_\s =\ha (A_+ - A_-)$ but not on $A_\tau =\ha (A_+ + A_-)$.
Hence, we may replace $A_\s$ by its on shell value using \eqn{apam}.
%
%
The result is action \eqn{smoac} with couplings \eqn{coupl}.
Needless to say that if we use the second
decomposition $l=h \tilde g$ then we obtain the dual $\s$--model
action with couplings \eqn{coupldu}.

It is instructive to see the meaning of the
canonical transformation \eqn{cantr} in this formalism. 
We evaluate
$\langle{l\inv \del_\pm l|\tilde T^a\rangle}$ in the alternative ways
suggested by the two possible parametrizations of $l\in D$
\ba
\langle{l\inv \del_\pm l|\tilde T \rangle}\big |_{A_\pm}^{l=\tilde h g} & = &
i (E_0^\mp)\inv E^\mp (L_\pm \pm \Pi F^\mp \del_\pm Y) ~ , 
\label{cccaa1}\\
\langle{l\inv \del_\pm l|\tilde T \rangle}
\big |_{\tilde A_\pm}^{l=h \tilde g}  & = &
\pm i \tilde E^\mp (\tilde L_\pm \mp F^\mp \del_\pm Y) ~ ,
\label{cccaa}
\ea
where as indicated, we have accordingly substituted 
for $A_\pm$ or $\tilde A_\pm$.
Then let us consider the transformation between variables of the two dual 
models defined as
\be
\langle{l\inv \del_\pm l|\tilde T^a\rangle}\big |_{A_\pm}^{l=\tilde h g} = 
\langle{l\inv \del_\pm l|\tilde T^a\rangle}
\big |_{\tilde A_\pm}^{l=h \tilde g}  ~ .
\label{canoll}
\ee
It is a straightforward computation to show that 
this\footnote{Using \eqn{rree} and 
the equations of 
motion \eqn{exil} we may cast \eqn{canoll} into the equivalent form 
\be
\langle{l\inv \del_\pm l| R^\pm_a\rangle}\big |_{A_\pm}^{l=\tilde h g} = 
\langle{l\inv \del_\pm l| R^\pm_a\rangle}
\big |_{\tilde A_\pm}^{l=h \tilde g}  ~ .
\label{canll}
\ee
}
implies the canonical transformation \eqn{cantr}, \eqn{cantr1}.
Vice versa, it can be shown that similarly to 
the cases of Abelian and non--Abelian duality \cite{sfeleuv,KSsusynab},
the transformations \eqn{cantr1}, \eqn{cantr} and the requirement 
for 2--dim Lorentz
invariance on the world--sheet implies \eqn{canoll} as well as \eqn{coupl}
and \eqn{coupldu}. 

At this point we shall very briefly mention an alternative formulation of 
Poisson--Lie T--duality in the Drinfeld double \cite{TyuUng}.
In this construction the WZW action $I_0(l)$ is also involved but in
contrast to \eqn{actiL} it is defined with the usual light--cone
variables $\s^\pm$. Also in contrast to \eqn{actiL} the 
entire action of \cite{TyuUng} is manifestly 2--dim Lorentz invariant.
However, in order to recover the dual $\s$--model actions 
one of the equations in \eqn{exil} is imposed as a constraint. 
A clarification of the relation between the two constructions is on open
problem and it might help to find the off shell formulation 
of supersymmetric \PL\ as we shall soon argue.

\section{Supersymmetric Poisson--Lie T--duality}

Extending our previous discussion to $N=1$ supersymmetric models 
is not entirely trivial. One approach is to start 
with the supersymmetric
versions of the bosonic equations of motion \eqn{exil}. They are given by
\be
\langle{L\inv D_\pm L |R^\mp_a\rangle}  = \pm i D_\pm Z^i F^\pm_{ia}~ . 
\label{sexil}
\ee
where $L$ and $Z^i$ are superfields corresponding to the
bosonic group element $l\in D$ and the spectator field $Y^i$ and
of course $F^\pm_{ia}$ depends on the $Z^i$'s.
The superfield  $L$ and its inverse have expansions in terms of 
anticommuting Grassman variables $\th_\pm$ given
by (see for instance \cite{veckni})
\ba
L & = & l - i \th_+ \chi_- l + i \th_- l \chi_+ - i \th_+ \th_- f ~ , 
\nonumber \\
L\inv & = & l\inv  + i \th_+ l\inv \chi_-  
- i \th_-  \chi_+ l\inv  -i \th_+ \th_-  f^+ ~ ,  \label{sexp} \nonumber \\
f^+ & = & - l\inv f l\inv + i (\chi_+ l\inv \chi_- 
- l\inv \chi_- l \chi_+ l\inv) ~ ,
\label{ssup} 
\ea
where $\chi_\pm \in \cD$ are world--sheet fermions and $f$ is the
highest component of the superfield. 
Also the world--sheet superderivatives are defined 
as $ D_\pm = \mp i \del_{\th_\mp} \mp \th_{\mp} \del_\pm$.
For the superfield $Z^i$ an analogous expansion holds 
\be
Z^i = Y^i - i \th_+ \Psi^i_- + i \th_- \Psi^i_+ - i \th_+ \th_- F^i~  .
\label{sspec}
\ee
Similarly to the bosonic case, we may use two alternative decompositions
$L=\tilde H G$ or $L= H \tilde G$, where the superfields $H,G\in G$ and 
$\tilde H, \tilde G \in \tilde G$ have a similar to \eqn{ssup} 
expansion
in terms of Grassman variables. Then after solving for the 
corresponding ``gauge fields'' it is obvious that the 
$N=1$ dual supersymmetric $\s$--models for the superfields $(G,Z^i)$ and 
$(\tilde G,Z^i)$ are given by the straightforward $N=1$ supersymmetrized
bosonic actions, namely by replacing 
the various fields and derivatives by their supersymmetric counterparts.

We have just presented an on shell formulation of supersymmetric
Poisson--Lie T--duality based on equations \eqn{sexil}. However,
an off shell formulation is not as straightforward as one might expect
since it requires an $N=1$ supersymmetric action with a local invariance
similar to the one for the bosonic action \eqn{actiL}
which then can be used 
to cast the corresponding equations of motion into the
form \eqn{sexil}. Although the supersymmetrization of the WZW action $I_0(l)$
in \eqn{actiL} is known \cite{veckni}, there are certain technical 
difficulties 
associated with the other terms. A possible way around this problem 
could be to
use the alternative formulation \cite{TyuUng} of \PL\ in the Drinfeld double 
we have mentioned. 
Since it is manifestly 2--dim Lorentz invariant the 
supersymmetrization of the action and the constraint causes no problem. 
Hence, finding the
precise relationship between the two formulations of \PL\ in 
\cite{TyuUng} and \cite{DriDou} seems important also in a
supersymmetric context.
We hope to report progress along these lines in the future \cite{klisfe}.

Next we turn into the behaviour of extended $N\geq 2$ supersymmetry 
under \PL. 
A complete discussion at the level of (super)Drinfeld double
requires knowledge of the $N=1$ supersymmetric version of the bosonic
action \eqn{actiL} which as mentioned we lack. Nevertheless, we may still 
discuss the situation at the level of the  
$\s$--models whose bosonic parts are given by \eqn{smoac} with couplings 
\eqn{coupl} and by its dual action with couplings \eqn{coupldu}.
Hence, we are looking for (generally non--local) complex structures satisfying
\eqn{nocos}. Quite generally they have the form 
\ba
F^\pm & = &  C^\pm_{AB}(Y) f^A_\pm \wedge f^B_\pm \nonumber \\
&  = & C^\pm_{ab} f^a_\pm \wedge f^b_\pm 
+ 2 C^\pm_{ia} \del_\pm Y^i \wedge f^a_\pm
+ C^\pm_{ij} \del_\pm Y^i \wedge \del_\pm Y^j ~ ,
\label{comform}
\ea
where $f_+^a$ and $f_-^a$ are some (1,0) and (0,1) forms on the world--sheet.
A similar to \eqn{comform} expression holds for $\tilde F^\pm$ as well.
Various interesting cases are:

\no
\underline{case (i)}: The $f^a_\pm$'s have the form 
\be
f^a_\pm = \langle{l\inv \del_\pm l|\tilde T^a\rangle}
\big |_{ A_\pm}^{l=\tilde h  g} ~ ,
\label{a1}
\ee
where the right hand side is given explicitly by \eqn{cccaa1}.
In the dual model $\tilde f^a_\pm$ is given instead by \eqn{cccaa}.
Hence, in this case extended supersymmetry is local in both dual
models even though the dual complex structures are not invariant under 
left(and right) transformations of $g\in G$.

\no
\underline{case (ii)}: The $f^a_\pm$'s are simply given by
\be
f^a_\pm = L^a_\pm ~ .
\label{a2}
\ee
In the dual model $L^a_\pm$ is expressed using \eqn{cccaa1}--\eqn{canoll} as
\be 
L^A_\pm = Q^A_{\pm B} \tilde L^B_\pm ~ ,
\label{lql}
\ee
where 
\be
Q_\pm =
\pmatrix{\pm (E^\mp)\inv E^\mp_0 \tilde E^\mp & - (I-\Pi \tilde \Pi) \tilde 
E^\mp F^\mp \cr
& \cr
0 & I } ~ .
\label{Apm}
\ee
In order to find the precise mapping under \PL\ 
one has to solve \eqn{lql} for $g\in G$. We haven't been able to 
do so but it is certain that it 
will be a non--local functional of the dual 
model target space variables. Hence, the corresponding extended supersymmetry
will be non--locally realized in the dual model, even though the original 
complex structure \eqn{comform}, \eqn{a2} was invariant under the 
left action of the group $G$. The non--local complex structure will be of
the form \eqn{comform} but with $\tilde C^\pm_{AB}$ given by
\be 
\tilde C^\pm_{AB} = Q^C_{\pm A} Q^D_{\pm B} C^\pm_{CD}~ .
\label{tilc}
\ee

\no
\underline{case (iii)}: In this case we consider
simultaneously more than one complex structures of the particular form 
\be 
(F^\pm)^a = \langle{l\inv \tilde T^a l| T_b\rangle}
\big |_{\tilde A_\pm}^{l=h \tilde g} (F^\pm_0)^b ~ ,
\label{fpm0}
\ee
where $(F^\pm_0)^a$'s are 2--forms similar to \eqn{comform} with
$f_\pm^a$'s given as in case (i) by \eqn{a1}.
We compute 
\be
\langle{l\inv \tilde T l| T \rangle}
\big |_{A_\pm}^{l=\tilde h  g} = \tilde a(\tilde h) 
(a^T(g))\inv ~ , ~~~~~
\tilde h = P e^{\int^\s  A_\s} ~ ,
\label{iii}
\ee
where $P$ stands for path ordering and 
$A_\s\equiv \ha(A_+ - A_-)$ is found using \eqn{apam}.
In the dual decomposition $l=h\tilde g$, the 2--form
$(F^\pm_0)^a$ remains local in the dual space target space variables as well,
but analogously to \eqn{iii} we have 
\be 
\langle{l\inv \tilde T l| T \rangle}
\big |_{\tilde A_\pm}^{l=h \tilde g} = b(h) \tilde b(\tilde g)
+(a^T(h))\inv \tilde a(\tilde g) ~ ,~~~~~
h = P e^{\int^\s \tilde A_\s} ~ .
\label{iij}
\ee
Hence, in both models extended supersymmetry is realized non--locally
and this is a novel characteristic of \PL.
A similar conclusion is reached in a variation of case (iii),
namely when $f^a_\pm$ is given by \eqn{a2} instead of \eqn{a1}.


\section{Discussion and concluding remarks}

We have reviewed aspects of \PL\ and explicitly formulated it as 
a canonical transformation on the string world--sheet.
We gave an on--shell formulation of supersymmetric $N=1$ \PL\ 
at the level of the two dual $\s$--models and examined
the behaviour of extended world--sheet supersymmetry under \PL.
An open interesting problem is to find the corresponding off--shell 
formulation fully at the level of the Drinfeld double.

This work is part of a program whose aim is to use duality 
symmetries in order to attack problems in fundamental Physics from a string 
theoretical point of view. An outstanding such problem is the
information loss paradox in black hole Physics via Hawking radiation.
A preliminary exploratory step in using \PL\ in this direction would be to 
search for 
non--trivial backgrounds, in the sense of \eqn{ftilf}, that are dual to 
black holes. In addition, discovering unconventional non--locally
realized supersymmetry in black holes or equivalently their duals
might stabilize them against 
Hawking radiation similarly to the case of 
extremally charged 
black holes whose stability may be attributed
to their supersymmetric properties. 
We hope to be able to report work in this direction in the future.

\newpage

\centerline{\bf Acknowledgments }
\no
I would like to thank the organizers of the Buckow Symposium for the 
invitation to present this work, for financial support as well as for 
their warm hospitality during the symposium.
I also thank C. Klimcik for useful discussions, the Argonne Nat. Lab.
for hospitality and financial support at the initial stage of this research
and the Physics Department of the University of Patras for hospitality
at the final stages of typing this paper.  
This work was also carried out with the financial support 
of the European Union Research Program 
``Training and Mobility of Researchers'', under contract ERBFMBICT950362.
It is also work supported by the European Commision TMR program
ERBFMRX-CT96-0045.



\end{document}